# Electro-Optical Sampling of Single-Cycle THz Fields with Single-Photon Detectors

Taylor Shields[1,*], Adetunmise C. Dada[2], Lennart Hirsch[1], Seungjin Yoon[1], Jonathan M. R. Weaver[1], Daniele Faccio[2], Lucia Caspani[3], Marco Peccianti[4], Matteo Clerici[1,**]

[1]James Watt School of Engineering, University of Glasgow, Glasgow G12 8QQ, United Kingdom
[2]School of Physics and Astronomy, University of Glasgow, Glasgow G12 8QQ, United Kingdom
[3]Institute of Photonics, Department of Physics, University of Strathclyde, Glasgow G1 1RD, United Kingdom
[4]Emergent Photonics Research Centre and Dept. of Physics, Loughborough University, Loughborough, LE11 3TU, UK
[*]t.shields.1@research.gla.ac.uk, [**]matteo.clerici@glasgow.ac.uk

**Abstract**

Electro-optical sampling of Terahertz fields with ultrashort pulsed probes is a well-established approach for directly measuring the electric field of THz radiation. This technique usually relies on balanced detection to record the optical phase shift brought by THz-induced birefringence. The sensitivity of electro-optical sampling is, therefore, limited by the shot noise of the probe pulse, and improvements could be achieved using quantum metrology approaches using, e.g., NOON states for Heisenberg-limited phase estimation. We report on our experiments on THz electro-optical sampling using single-photon detectors and a weak squeezed vacuum field as the optical probe. Our approach achieves field sensitivity limited by the probe state statistical properties using phase-locked single-photon detectors and paves the way for further studies targeting quantum-enhanced THz sensing.

## Introduction

Terahertz radiation is traditionally defined as the range of the electromagnetic spectrum between the microwave and infrared regions and in recent years has garnered significant interest in fields such as imaging, medicine, communications, etc. [1–5]. A wide range of applications of THz radiation relies on THz time-domain spectroscopy (THz-TDS), a technique able to measure THz electric fields. This uses ultrafast laser pulses to generate and subsequently measure the THz radiation in the time domain exploiting nonlinear optical processes. A well-established THz-TDS system approach is based on electro-optic sampling, a form of time-resolved polarimetry that utilizes the Pockels Effect in second-order nonlinear crystals to induce a phase shift proportional to the THz field strength [1–3,6–9]. This way, the THz radiation can be reconstructed in the time domain if the probe pulse is shorter than the THz electric field oscillation period. Currently, the sensitivity of THz-TDS is limited by the shot noise of the optical probes. Remarkably, under appropriate experimental conditions, it has been recently shown that vacuum fluctuations and intensity correlations can be observed by electro-optical sampling at infrared and THz frequencies [10–14]. Efforts to further enhance performance have so far mainly targeted the optimization of the detection approach [15].

THz-TDS is, in essence, a polarimetric measurement, and its sensitivity could therefore be improved by applying quantum metrology strategies aimed at overcoming the shot-noise limit in polarization and phase sensing [16], as also recently theoretically proposed in [17]. The discrete variable quantum metrology approaches conventionally applied to polarimetry exploit quantum resources such as NOON [18] and Fock states [19,20]. They rely on single-photon detectors and squeezed vacuum states (photon pairs) generated by parametric down-conversion (PDC) in second-order nonlinear crystals. Therefore, a first step towards implementing quantum metrology to enhance THz-TDS requires investigating under what experimental conditions electro-optical sampling can be performed with single-photon detectors.

In this work, we have investigated this aspect, performing electro-optical sampling of THz waves using as a probe ultrashort squeezed vacuum radiation generated by the PDC of a 100 fs duration optical pulse at 780 nm in a periodically poled lithium niobate crystal. To this end, we developed a scheme to perform a lock-in-style polarimetry with single-photon detectors, achieving detection with a sensitivity limited by the state statistics, thus paving the way for the use of nonclassical probes in THz electro-optical sampling aiming toward quantum-enhanced time-domain spectroscopy. We used squeezed vacuum, i.e., the radiation generated by spontaneous PDC, rather than a weak coherent probe pulse, as the former is a widely employed resource underpinning quantum-enhanced metrology.

**Standard THz-TDS measurement**

As a first step, we generated a broadband (single-cycle) THz electric field and we measured it in a standard THz-TDS setup, as described below. The THz radiation is generated by a photoconductive antenna (Tera-SED, Laser Quantum, Stockport, UK) photo-excited with a train of ultrashort pulses delivered by a mode-locked laser (Coherent Discovery) with an average power of 600 mW, pulse duration of 100 fs, 80 MHz repetition rate, and central wavelength of 780 nm. The THz antenna consists of a planar $3 \times 3$ mm$^2$ GaAs substrate with an interdigitated metal-semiconductor-metal electrode structure, which is modulated by an externally applied bias field.

The employed photoconductive antenna (PCA) delivers a single-cycle THz field with a peak emission frequency of 1.5 THz. The excitation condition was optimized by recording the THz power with a pyroelectric detector (THZ5I-BL-BNC-D0, Gentec EO, Québec city, Canada) aided by a low noise current mode amplifier and using a 5 Hz chopping frequency [21,22]. The most effective spot size at the maximum excitation power of 600 mW was measured to be 297 μm. The THz electric field trace is then measured using a conventional THz-TDS setup. In our condition, the birefringence of a 250 μm thick AR-coated (1560 nm) GaAs <110> crystal is modulated by the THz electric field and probed by polarimetry performed with a short optical pulse overlapped to different temporal sections of the THz field [23]. The residual pump field incident on the PCA is removed with a paper filter before reaching the detection crystal. The 200 fs probe pulse used for the optical sampling is delivered by an Optical Parametric Oscillator (OPO, Levante IR, Coherent, Santa Clara, California, United States) tuned at 1560 nm, at the same wavelength of the squeezed vacuum source employed for the single-photon-level measurement. The relative delay between the probe and the THz pulses is controlled by a delay stage (M-VP-25XL, Newport, Irvine, California, United States).

The phase delay induced by the THz field on the probe pulse is analyzed by means of a standard polarimetric arrangement comprised of a quarter-wave plate (λ/4, Newport 10RP04-40) followed by a Wollaston prism (WPA10, Thorlabs, Newton, New Jersey, United States) and a balanced detector (PDB210C/M, Thorlabs). The THz and the probe field were cross-polarized and the orientation of the GaAs was optimized to maximize the electro-optical modulation [24]. The quarter-wave plate is oriented in such a way that no differential signal is produced by a probe injected into the setup in absence of the THz field. The balanced detector amplified signal is measured and digitalized by a lock-in amplifier (SR8300, Stanford Research

Systems, Sunnyvale, California, United States), synchronized to the 10 kHz signal employed to modulate (on-off) the 30 V bias voltage applied to the PCA. Using high lock-in frequencies allows for reducing the impact of the 1/*f* noise [25–27]. In Figure 1A we show an example of a THz trace acquired before (blue) and after (red) purging our setup with pure nitrogen to reduce atmospheric water absorption. In Figure 1B we show the spectra obtained by Fourier transform of the time domain signals in Figure 1A. The strong absorption features of atmospheric water are clearly evident [28,29]. The humidity of the experimental enclosure was continually monitored using an OMEGA iBTHX relative humidity sensor with 0.1% resolution.

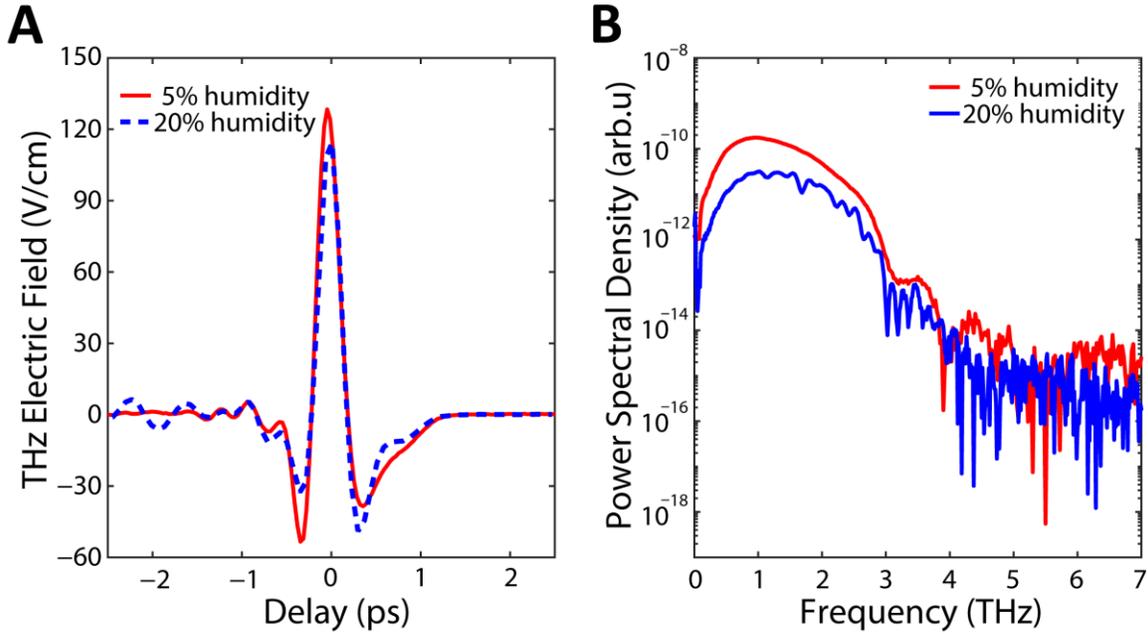

**Figure 1.** (**A**) THz time-domain spectroscopy using electro-optic sampling to characterize the THz field trace and spectrum using a classical balanced detection scheme. (**B**) Power spectrum corresponding to the measurement in (**A**). The blue curves are for an un-purged setup (blue curve) while the red curves show the result in a pure nitrogen atmosphere.

The amplitude $E_{THz}(t)$ of the measured THz electric field trace $E_{tr}(t)$ was calibrated according to [30]:

$$E_{THz}(t) = \frac{E_{tr}(t)}{max(|\mathcal{E}_{tr}(t)|)} \sqrt{\frac{2U\eta_0}{\int G(x,y)\, dx\, dy \int \frac{|\mathcal{E}_{tr}(t)|^2}{max(|\mathcal{E}_{tr}(t)|^2)} dt}} \quad (1)$$

where $\mathcal{E}(t)$ is the complex electric field trace, such that $E_{tr}(t) = Re[\mathcal{E}_{tr}(t)]$, $U \simeq 1.25$ fJ is the THz pulse energy measured with the pyroelectric detector, $G(x,y) = \exp(-x^2/\sigma_x^2 - y^2/\sigma_y^2)$ is the THz beam profile determined by knife-edge measurements with $\sigma_x = \sigma_y = 200 \pm 10$ µm, and $\eta_0$ is the vacuum impedance (see Supplementary Materials for details on the knife-edge measurement). According to this calibration, the THz field amplitude reached a peak value in air of

$\simeq 130$ V/cm, as shown in Figure 1A.

A crucial step required to perform the THz measurement with single-photon detectors is to assess the detection system sensitivity and the magnitude of the phase shift the generated THz radiation imparts on a probe field. To this end, we have estimated the maximum phase shift $\Delta\Phi \simeq \frac{\Delta I_p}{I_p}$ (relative difference in the probe signal intensity between the two photodiodes) that the THz field in our experiment will be able to impart to a probe field inside the detection crystal according to [24,31]:

$$\Delta\Phi = \left|\frac{E_{THz} r_{41} n^3 \omega L}{c}\right| \qquad (2)$$

where $r_{41}$, $n$ and $L$ are the detection crystal electro-optical coefficient, refractive index and thickness. Using the recorded THz field amplitude and considering $r_{41} \simeq 1.5$ pm/V, $n = 3.38$, and $L \simeq 250$ μm we estimated a maximum phase shift of $\Delta\Phi \simeq 2.4 \times 10^{-4}\, \pi$. The estimation of the maximum phase shift our peak field induces allowed us to evaluate the integration time required to observe the phase change with single-photon detectors.

**Generation of squeezed vacuum and measurement calibration**

Once the THz radiation properties were characterized with a standard THz-TDS setup using a coherent probe and proportional detectors, we set up the squeezed vacuum source to be used as an alternative probe pulse. The squeezed vacuum is a state of light with non-classical properties such as energy-time entanglement and is one of the most employed resources in quantum optics. It is routinely generated by spontaneous PDC in second-order nonlinear crystals [32] and has a thermal statistical distribution with only even photon numbers occupied, i.e., it consists of photon pairs (or higher even numbers of photons) [33]. In our experiment, squeezed vacuum with a wavelength centered at 1560 nm was generated by Type 0 PDC of the 780 nm pump laser in a 0.5 mm long periodically poled Magnesium-doped lithium niobate crystal (MgO-PPLN with 19.70 μm poling period, MSHG1550-0.5, Covesion, Southampton, United Kingdom). Magnesium Oxide doping is used to provide more flexibility for degenerate wavelength tuning and for the optical and photorefractive durability in comparison to undoped PPLN [34]. Off-axis parabolic mirrors of an equivalent focal length of 50 mm were used to focus the pump beam to 13 μm into the crystal and to collect and collimate the generated radiation.

We first characterized the source in the high photon-flux regime using an amplified InGaAs photodiode (Thorlabs PDA10DT) varying the crystal temperature to maximize the flux within a 12 nm spectral region centered at 1560 nm (FBH1550-12, Thorlabs). We then routed the squeezed vacuum radiation into the balanced detection setup employed for the characterization of the THz-induced phase shift. Such a setup mimics an ellipsometry measurement of standard EOS. The squeezed vacuum probe field is split into two polarization components using a quarter-wave plate (QWP, 10RP04-40, Newport), half-wave plate (HWP, B. Halle RAC 6.2.15) and a polarizing beam splitter (PBS, Thorlabs PBS204). Differently from before, however, the collimated output of the polarimetric setup is coupled into optical fibers using reflective collimator couplers (RC04FC-P01, Thorlabs). A lens telescope ($f_1 = -75$ mm, $f_2 = +150$ mm) was also inserted into the beam path to optimize the coupling. The radiation was then measured with InGaAs-based Single-Photon Avalanche Photodiodes (ID Quantique, ID230, Geneva, Switzerland). These detectors can record the arrivals of photons at a rate limited by a dead time as short as 2 μs and with 25% quantum efficiency. The arrival time of the photons was digitalized with time-tagging electronics (PicoQuant, HydraHarp, Berlin, Germany).

To calibrate our single-photon-detector-based polarimetry setup and to assess its phase sensitivity, we used an electro-optical modulator (EOM—Thorlabs, EO-AM-NR-C3), as shown in Figure 2A, configured to change the polarization of the injected radiation. The half-wave voltage ($V_\pi$) at the operating wavelength was 539 V, and the EOM was driven by a high-voltage amplifier (max 200 V, HVA200, Thorlabs) fed with a modulated signal.

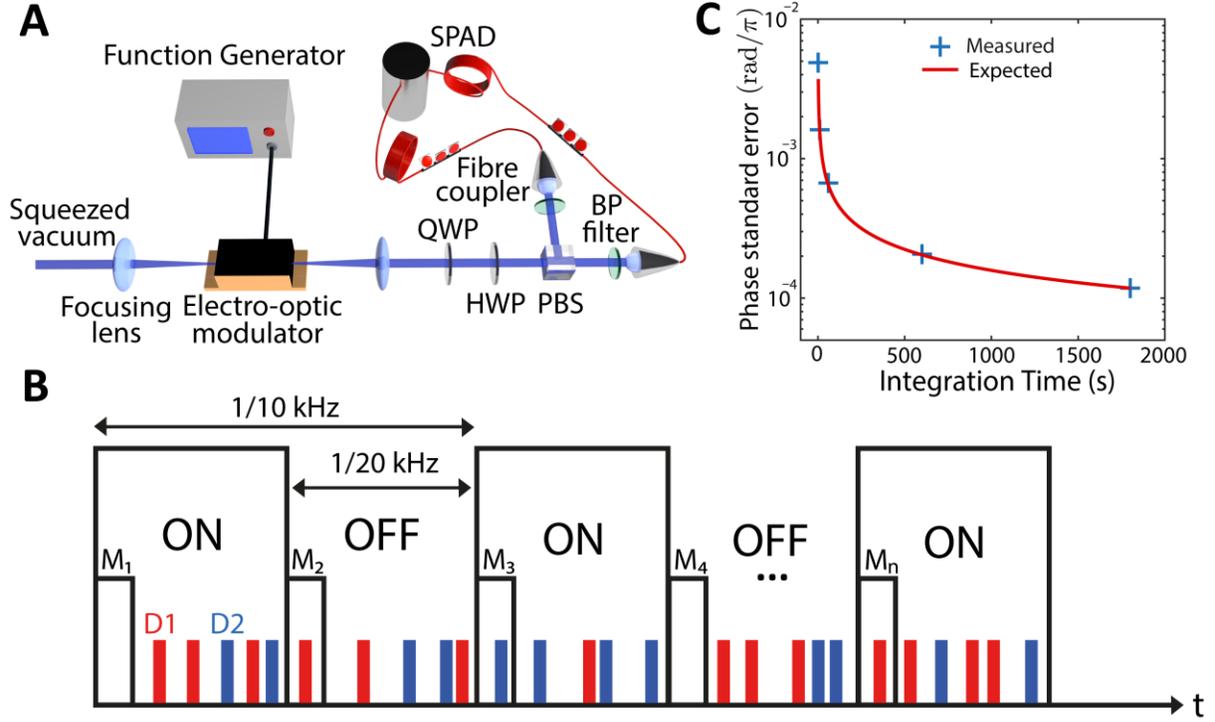

**Figure 2.** Calibration setup with EOM and experimental lock-in-like balanced detection using two single-photon detectors. (**A**) Experimental scheme for calibration of detection using electro-optic modulation. (**B**) Using ps-time tagging resolution of the photon counting software, a difference measurement was performed between counts of detectors D1 (red) and D2 (blue) using markers M inserted every 1/20 kHz to synchronize timing with the antenna modulation. (**C**) Measured standard error of the detection method using EOM with single photon counts.

In an ideal case, the phase shift induced by the EOM will result in an unbalance of the signal on one photon counter with respect to the other, such that $\Delta N \simeq (N_1 + N_2)\Delta\Phi$, where $N_1$ and $N_2$ are the total number of photons on the two channels. To eliminate the effect of technical noise, we have implemented a digital analogue of lock-in detection shown in Figure 2B, performed by modulating the EOM using a square wave with a $f = 10$ kHz frequency and introducing markers in the time-tagged data at double such frequency. We note that the modulation frequency is the same used to drive the PCA for the THz measurement described above. For each modulation period $l$ comprised of $M = 80 \text{ MHz}/(2 \times f) = 4000$ laser pulses, we recorded a phase shift:

$$\Delta_r \Phi_l = \langle \Delta_r \Phi_{\text{on}} \rangle_l - \langle \Delta_r \Phi_{\text{off}} \rangle_l \qquad (3)$$

where $r$ stands for "recorded", on and off refer to the sections of the time-tagged data where the modulation was active or inactive, respectively, and $\langle \cdot \rangle$ is an average over $M$ laser pulses.

However, single-photon detectors are not ideal photon counters and return one count also in cases where more than one photon is impinging at the same time, and for this reason, are referred to as on/off detectors [35]. A relation between the effective photon flux and the measured one can be derived considering the statistical properties of the state employed in the measurement and losses (see Supplementary Materials for details). The effective phase shift can then be assessed as

$$\Delta\Phi_l = \alpha(n_{sv}, \eta)\Delta_r\Phi_l \tag{4}$$

where $\alpha(n_{sv}, \eta)$ includes the effects of using on/off detectors and losses and is estimated as discussed in the Supplementary Materials and $n_{sv}$ is the average photon number of the squeezed vacuum probe. The value of the phase shift $\Delta\Phi$ is then obtained by fitting the $\Delta\Phi_l$ distribution sampled performing $K = T \times f$ repeated measurements, with $T$ the integration time. The standard error of the effective phase, $\sigma_K(\Delta\Phi)$, is decreasing as expected with $K^{-1/2}$ as shown in Figure 2C.

## THz field measurement with single-photon detectors

Once proven that the detection approach works in the simpler case of electro-optical modulator, we substituted the EOM with the GaAs detection crystal excited with the THz radiation, purged the generation and detection area with nitrogen, and recorded the data as described above. A complete sketch of the experimental setup is shown in Figure 3A. Additionally, in this case, the temporal delay of the THz field relative to the probe pulse was controlled with a linear stage (M-VP-25XL, Newport) and scanned in an identical fashion to the THz-TDS measurement in Figure 1A. The estimated THz peak field measured with the classical THz-TDS setup ($\simeq 130$ V/cm) corresponds to a phase shift of $\Delta\Phi \simeq 2.4 \times 10^{-4}\pi$. To measure such a small phase shift we employed superconducting nanowire single photon detectors (Single Quantum, Delft, Netherlands) as they enable higher rates (up to MHz, not limited by the SPAD deadtime) and have efficiencies of $\eta = 0.65$ (measured in our experimental condition including channel losses). Figure 3B shows the measured phase shift induced by the THz radiation on the infrared squeezed vacuum probe pulse as a function of their relative delay (blue squares) compared against the THz trace measured with the classical THz-TDS obtained via electro-optic sampling (red curve, scaled). The error bars represent the standard error for each phase measurement and are calculated from the 68% confidence bounds of the fit performed on the experimental data. Note that, for this measurement, $\alpha \simeq 0.790$, see Equation (4) and Supplementary Materials. This measurement demonstrates that it is possible to perform THz-TDS with a probe pulse at the single-photon level (with an estimated average of 0.05 photons per pulse) and provides the first proof of an ultrafast phase modulation applied within a picosecond window on squeezed vacuum radiation. We estimated a phase shift of $\Delta\Phi \simeq 1.71 \times 10^{-4}\pi$, lower than the value measured with the intense classical probe ($2.4 \times 10^{-4}\pi$). We attributed this small difference to the shorter duration of the squeezed vacuum respect to the coherent pulse and residual spatial overlap issues. The THz electric field trace measured with squeezed vacuum is also shorter than the one recorded with a coherent probe. The difference can be attributed to the shortened duration of the squeezed vacuum (<100 fs) with respect to the coherent pulse (~200 fs).

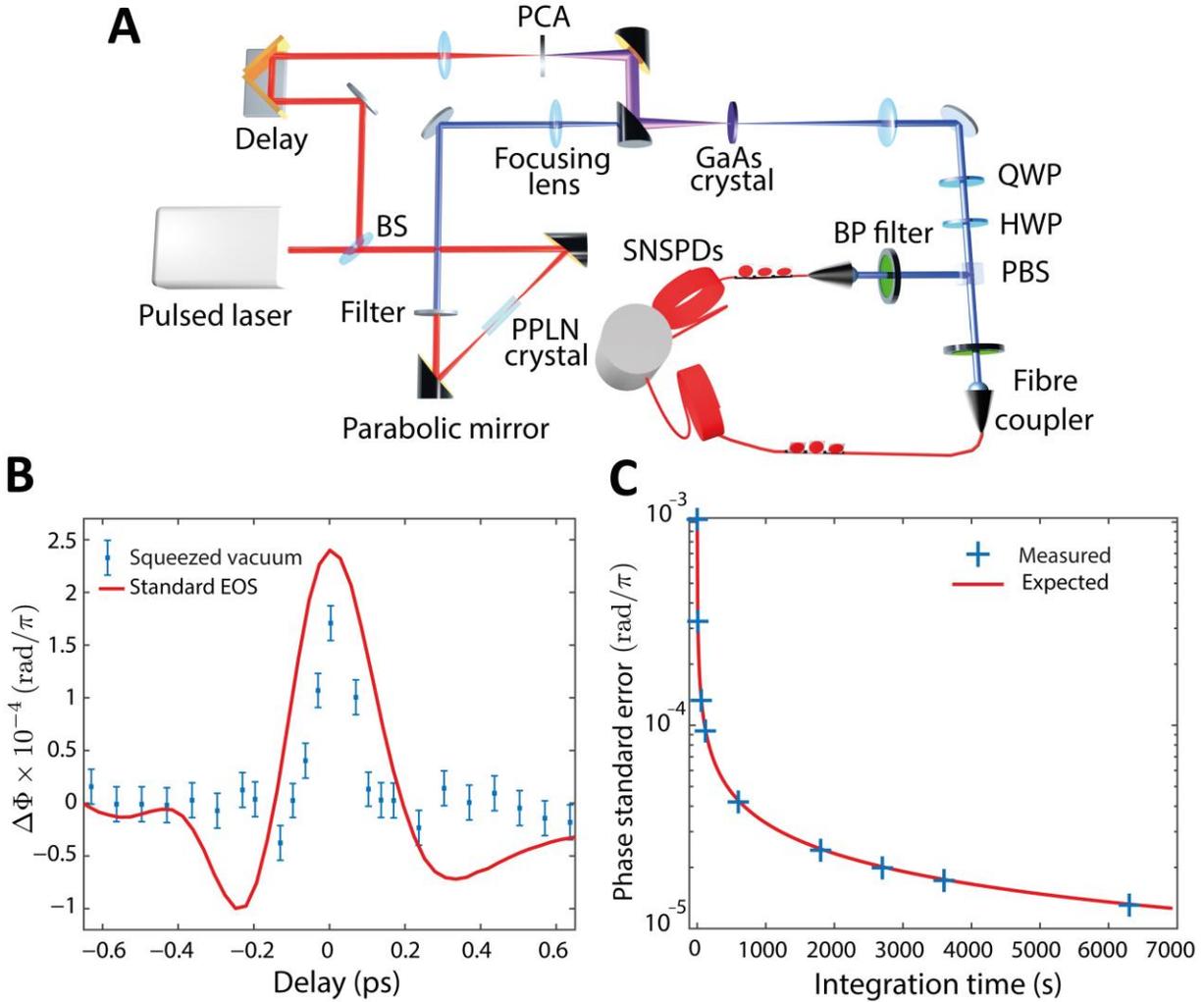

**Figure 3.** Terahertz detection using single photon detectors. (**A**) Experimental scheme for measuring THz field using squeezed vacuum. The THz field generated from a photoconductive antenna (PCA) is overlapped to the squeezed vacuum pulse into a GaAs electro-optical crystal. The phase shift introduced by the THz pulse on the squeezed vacuum probe is measured by a balanced detection using single photon detectors. (**B**) Measured phase shift ($\Delta\Phi$) induced by THz field (blue squares) overlapped with the phase shift from standard electro-optic sampling (red curve). The integration time for each data point was 105 min. (**C**) Measured standard error of the measurement (blue crosses) compared to the expected standard error calculated considering the squeezed vacuum probe statistics (red curve) at increasing integration time. The good match indicates that the detection sensitivity is limited by the statistical properties of the probe state, and the THz field is not adding noise.

We note that the sensitivity of the phase measurement is limited by the statistical properties of the employed state and the size of the acquired samples $K$. The expected standard error $\sigma_K(\Delta\Phi)$ can indeed be linked to the standard deviation $\sigma(n^-)$ of the distribution of expected counts from a differential measurement between the output ports of a 50:50 beam splitter upon injection of a single mode squeezed vacuum state according to:

$$\sigma_K(\Delta\Phi) = \frac{\sigma(\Delta\Phi_l)}{\sqrt{K}} \simeq \sqrt{\frac{1}{K}} \alpha\, \sigma(\Delta_r \Phi_l) \simeq \sqrt{\frac{2}{K\,M}} \alpha\, \frac{\sigma(n^-)}{n^+} \qquad (5)$$

where $n^+$ is the expected sum of the counts among the output ports of the 50:50 beam splitter. As discussed in the Supplementary Materials, for the parameters of the experiments, $\eta = 0.65$ and average photon number per pulse measured on each detector of 0.0135 (corresponding $\simeq$ 1 MHz rates per channel), we have $\alpha \simeq 0.790$ and $\sigma(n^-) = 0.126$. In Figure 3C we compare the standard error of the THz-induced phase at the THz peak temporal coordinate (blue crosses) with the expected values from Equation (5) (red curve). Note that $\sigma(n^-)$ is strongly dependent on the particular statistics of the probe state and the good match between the model and the measurement indicates that our THz-TDS is limited in sensitivity by the statistical properties of the employed state.

Finally, we note that the standard deviation of the $n^-$ distribution for a squeezed vacuum probe can be smaller than that for a coherent probe with the same average photon number using on/off detectors (see Supplementary Materials for details). However, the sensitivity of the polarimetric measurement in both cases is the same in absence of losses when employing number-resolved detectors, while the squeezed vacuum performs slightly worse than a coherent state if losses and on/off detectors are considered.

## Conclusions

Using the lock-in style phase approach, we have demonstrated time-resolved detection of THz single-cycle pulses employing a squeezed vacuum source and single-photon detectors. The acquisition setup sensitivity is limited by the probe field statistical properties and allowed the observation of a weak THz single-cycle pulse with peak amplitude of $\simeq$ 130 V/cm with a signal-to-noise ratio of $\simeq$ 10 (ratio between the measured peak phase and the mean standard error) for an acquisition time of 105 min. While no quantum enhancement in the THz detection can be proved at this stage, our work lays the basis to apply quantum-metrology strategies to enhance the detection sensitivity of THz fields, employing non-classical resources such as NOON states or single-photons as a probe light for the THz-TDS scheme. Furthermore, we have shown that sub-picosecond single-cycle THz pulses can impart a phase shift to independently detected photons, albeit limited in our case to ~0.05 mrad. As large THz peak fields can be achieved enabling full-wave phase shifts, we foresee a potential application of the scheme demonstrated here to ultrafast encoding of phase information [36] and polarization demultiplexing in single-photon applications.

**Supplementary Materials:** The following supporting information can be downloaded at: https://www.mdpi.com/article/10.3390/s22239432/s1, S.1. THz Knife-edge Measurement; S.2. Statistical properties of the squeezed vacuum; S.3. Detection sensitivity; S.4. Statistical properties of the state realized in the experiments; Supplementary References; Figure S1: Beam splitter geometry; Figure S2: Theoretical analysis of the standard deviation and sensitivity; Figure S3: Experimental data statistical properties; Figure S4: Theoretical analysis of the expected measurement sensitivity including losses.

**Author Contributions:** T.S., A.C.D., L.C. and M.C. designed the experiment. T.S. performed the measurements with contributions from A.C.D. and M.C. T.S., A.C.D., L.H., S.Y., L.C. and M.C. analyzed the data and interpreted the results. J.M.R.W., D.F., L.C., M.P. and M.C. provided technical support to the experiment, data analysis, and interpretation of the results. T.S., A.C.D., S.Y., L.C. and M.C. drafted the manuscript. M.C. supervised the research activity. All authors have read and agreed to the published version of the manuscript.

**Funding:** M.C. and A.C.D. acknowledge the support from the UK Research and Innovation (UKRI) and the UK Engineering and Physical Sciences Research Council (EPSRC) (Fellowship "In-Tempo" EP/S001573/1). MC and LH acknowledge the support from the Defence Science and Technology Laboratory (DSTL), DSTLX-1000144632. M.C., S.Y., L.C. acknowledge the support from Innovate UK, Application Number PN 10001572 (HiQuED). LC acknowledges financial support from the UK Engineering and Physical Sciences Research Council (grant EP/V062492/1). DF acknowledges financial support from the Royal Academy of Engineering Chairs in Emerging Technology Scheme and from the UK Engineering and Physical Sciences Research Council (grants EP/M01326X/1 and EP/R030081/1). This project has received funding from the European Research Council (ERC) under the European Union's Horizon 2020 research and innovation programme Grant agreement No. 725046.

**Data Availability Statement:** All data underpinning this publication are available at the following address: https://doi.org/10.5525/gla.researchdata.1330 (accessed on 1 December 2022).

**Conflicts of Interest:** The authors declare no conflict of interest.